\documentclass[11pt,a4paper]{article}
\pdfoutput=1

\usepackage{jcappub}

\usepackage{amsmath}
\usepackage[normalem]{ulem}
\usepackage{graphicx}
\DeclareGraphicsExtensions{.eps, .eps, .jpg, .png}

\def\bk{{\bf k}}
\def\bq{{\bf q}}

\def\la{~\mbox{\raisebox{-.6ex}{$\stackrel{<}{\sim}$}}~}
\def\ga{~\mbox{\raisebox{-.6ex}{$\stackrel{>}{\sim}$}}~}

\title{An inflationary model with small scalar and large tensor nongaussianities}

\author[a]{Jessica L. Cook,}

\author[a]{Lorenzo Sorbo}

\affiliation[a]{Department of Physics, University of Massachusetts, Amherst, MA 01003}
\emailAdd{jlcook@physics.umass.edu, sorbo@physics.umass.edu}

\abstract{
We study a model of inflation where the scalar perturbations are almost gaussian while there is sizable (equilateral) nongaussianity in the tensor sector. In this model, a rolling pseudoscalar gravitationally coupled to the inflaton amplifies the vacuum fluctuations of a vector field. The vector sources both scalar and tensor metric perturbations. Both kinds of perturbations are nongaussian, but, due to helicity conservation, the tensors have a larger amplitude, so that nongaussianity in the scalar perturbations is negligible. Moreover, the tensors produced this way are chiral. We study, in the flat sky approximation, how constraints on tensor nongaussianities affect the detectability of parity violation in the Cosmic Microwave Background. We expect the model to feature interesting patterns on nongaussianities in the polarization spectra of the CMB. }

\begin{document}
\maketitle


\section{Introduction}%

The simplest models of inflation  predict a quasi-scale invariant spectrum of essentially gaussian scalar and tensor perturbations. The measurements of the Cosmic Microwave Background radiation are in wonderful agreement with these predictions (with the only caveat that we have not yet seen any evidence of tensor fluctuations). While the recent years have seen a significant amount of work on a possible nongaussian component in the spectrum of scalar perturbations generated during inflation, the possibility of nongaussianities in the tensor sector has raised less interest.  This is due to the fact that {\em (i)} tensor perturbations have a smaller amplitude than scalar ones by a factor $\sqrt{r}\lesssim .3$ and that {\em (ii)} by playing with the interactions of a scalar inflaton, it is in general easier to generate nongaussianities in the scalar perturbations than in the tensors. Literature  on tensor nongaussianities of inflationary origin includes~\cite{Maldacena:2011nz,Soda:2011am,Shiraishi:2011st,Gao:2011vs,Huang:2013epa,Zhu:2013fja} and focuses on models where the gravitational sector is suitably modified and/or directly coupled to the inflaton sector. Parity violating tensor nongaussianities produced after inflation by magnetic fields were studied in~\cite{Shiraishi:2012sn}.

In the present paper we discuss (to our knowledge, for the first time) a model where the scalar perturbations are gaussian to a very good approximation, whereas the tensor fluctuations have a larger and observable degree of nongaussianity. This implies that the nongaussianities in the three-point function of the temperature fluctuations will be essentially induced only by the tensor fluctuations.  Moreover, the tensor modes in the model we consider are essentially chiral. As a consequence: {\em (i)} since tensors contribute significantly to temperature fluctuations only at large scales, the nongaussian signal will be visible in the $\delta T/T$ maps only for moderately small multipole moments $\ell\lesssim 100$; {\em (ii)} we expect that nongaussianities in the polarization sector will have a particular form that might be detectable in polarization data, and {\em (iii)} chirality of the tensor modes will give a peculiar form to the three-point function (in particular, as discussed in~\cite{Kamionkowski:2010rb,Shiraishi:2011st} the coefficients $B_{\ell_1\ell_2\ell_3}$ will vanish for $\ell_1+\ell_2+\ell_3=$even, opposite to the case of nongaussianities sourced by scalar or parity-even tensors). 

The system we consider is given by an axion-like field $\chi$ interacting with a $U(1)$ gauge field via the operator 
\begin{equation}\label{int}
\frac{\chi}{f}\,F_{\mu\nu}\,\tilde{F}^{\mu\nu}\,,
\end{equation}
where $1/f$ is a coupling constant with the dimension of a length. This interaction is especially interesting because, when the zero mode of $\chi$  is evolving in time, the term~(\ref{int}) does not have a well-defined sign, and photons of one helicity see their mode functions exponentially amplified with amplitude $\approx \exp\{\dot{\chi}/f\,H\}$~\cite{Anber:2006xt}. These vectors interact with the inflaton and with gravity to lead to several interesting observational effects: nongaussianities in the scalar sector~\cite{Barnaby:2010vf,Barnaby:2011vw}, a running spectral index~\cite{Meerburg:2012id}, chiral tensor modes~\cite{Sorbo:2011rz}, gravitational waves that, at short wavelength, are so large to be detectable by the upcoming ground-based gravitational wave detectors~\cite{Cook:2011hg,Barnaby:2011qe}, primordial black holes~\cite{Linde:2012bt}. The possibility that the coupling of eq.~(\ref{int}) sources an anisotropic component of the power spectrum was discussed in~\cite{Dimopoulos:2012av,Urban:2013spa}.

In the simplest scenario where the field $\chi$ appearing in eq.~(\ref{int}) is the inflaton, non-observation of nongaussianities and of a running spectral index in the scalar perturbation sector imposes the strongest constraint on the coupling $1/f$, and as a consequence the amplitude of the tensor modes induced by the very same gauge fields at CMB scales is small and undetectable.  There are several ways out of this conclusion. In~\cite{Sorbo:2011rz} it was suggested that either the presence of a large number of gauge fields or the existence of a curvaton can effectively suppress the amplitude of the photon-induced scalar fluctuations and lead to an observable spectrum of chiral tensors. More recently, the paper~\cite{Barnaby:2012xt} has considered the possibility that the field $\chi$ is not the inflaton, but some other rolling pseudoscalar that is only gravitationally coupled to the inflaton. In this case, the nongaussian scalar perturbations are induced only though gravitational interactions and are suppressed with respect to the case where $\chi$ is the inflaton. As a consequence, nongaussianities in the scalar perturbations are small whereas chiral gravitational waves of a detectable amplitude can be generated. 

In the present paper we study the amplitude of nongaussianities in the {\em tensor} sector of the model of~\cite{Barnaby:2012xt}. As we will see the three-point function of the gravitational waves turns out to be three orders of magnitude larger than that of the scalars. The reason is that in this model both scalar and tensors are produced with comparable (gravitational) efficiency, but scalars pay an additional suppression due to helicity conservation. Since this mechanism is associated to sub-horizon dynamics, the shape of nongaussianities is equilateral.

We expect nongaussian tensors to significantly affect the three-point function of the $E$ and $B$ polarization modes. In this paper we do not consider this effect, leaving the analysis for future work, and we focus on the effect of nongaussian tensors on the temperature fluctuations.  We estimate the temperature bispectrum in the flat sky approximation and find that the Planck 2013 limits on $f_{NL}^{\rm {equil}}$ give, in most of the parameter space, the strongest constraints on the parameters of the model (note however that tensors affect temperature fluctuations only at large scales, $\ell\lesssim 100$, giving a strongly scale-dependent $f_{NL}^{\rm {equil}}$, so that the actual constraints on the model should be weaker than those derived from the limit on the scale invariant $f_{NL}^{\rm {equal}}$ discussed in~\cite{Planck2}). Nevertheless, if $r\lesssim 0.06$, tensors with a ${\cal O}(1)$ net chirality are compatible with the current constraints from nongaussianities. In this case, tensor chirality might be detectable, through the generation of non vanishing $\langle TB\rangle$ and $\langle EB\rangle$ correlators, in a CMBPol-like experiment~\cite{Saito:2007kt,Gluscevic:2010vv}. 

The plan of the paper is the following. In section II we review the mechanism by which a rolling axion-like field, through the amplification of a vector vacuum fluctuations, can source scalar and (chiral) tensor metric perturbations. In section III we compute the two- and three-point functions of the scalar and the tensors, showing that the tensor three-point function is a factor $\sim 10^3$ larger than the scalar one. In section IV we review the formalism of the flat sky approximation and apply it to temperature fluctuations sourced by tensors. Then we use this formalism to compute the scalar and tensor contribution to the bispectrum of the temperature perturbations. Finally, in sections V and VI, we discuss our results and present our conclusions.

\section{A rolling axion amplifying vectors, that generate scalars and tensors}

We consider the system described by the following lagangian
\begin{equation}\label{lagr}
{\cal L}=-\frac{1}{2}(\partial\phi)^2-V(\phi)-\frac{1}{2}(\partial\chi)^2-U(\chi)-\frac{1}{4}F_{\mu\nu}\,F^{\mu\nu}-\frac{\chi}{4\,f}F_{\mu\nu}\,\tilde{F}^{\mu\nu}\,,
\end{equation}
where the field $\phi$ is the inflaton and $\chi$ is a second (pseudoscalar) field that is rolling during inflation due to its potential $U(\chi)$. We will not be concerned about the specific dynamics of these fields, and we will simply assume that $V(\phi)$ can support inflation and that $\dot\chi$ is approximately constant during the period of generation of the perturbations relevant to the CMB.

\subsection{Rolling pseudoscalar amplifies vectors}

The last term in the lagrangian~(\ref{lagr}) is responsible for the amplification of the vacuum fluctuations of the gauge field. To analyze this phenomenon, we define the vector potential $A_\mu$ from $F_{\mu\nu}=\nabla_\mu A_\nu-\nabla_\nu A_\mu$ and we choose the Coulomb gauge $A_0=\partial_i\,A_i=0$. Then, neglecting the spatial gradients of $\chi$, the equation for the gauge field reads
\begin{eqnarray}
\label{a15}
A_i''-\Delta\,A_i-\frac{\chi'}{f}\,\epsilon_{ijk}\,\partial_j\,A_k=0\,,
\end{eqnarray} 
where the prime denotes differentiation with respect to the conformal time $\tau$ and $a(\tau)=-1/(H\,\tau)$ is the scale factor of the spatially flat, inflating Universe with Hubble parameter $H$. To study the amplification of the mode functions of the gauge field we promote the classical field $A_i(\tau,\,{\bf x})$ to an operator $\hat  A_i\left(\tau,\,{\bf x}\right)$, that we decompose into annihilation and creation operators $\hat{a}_\lambda^\bk$, $\hat{a}_\lambda^\bk{}^\dagger$
\begin{eqnarray}\label{a16}
{\hat A}_i(\tau,\,{\bf x})=\int\frac{d^3{\bf k}}{\left(2\pi \right)^{3/2}}e^{i{\bf k\cdot x}}{\hat A}_i(\tau,{\bf k})=\sum_{\lambda=\pm}\int \frac{d^3\bk}{\left(2\pi \right)^{3/2}}\left[\epsilon^i_\lambda(\bk)\,A_\lambda(\tau,\,\bk)\,{\hat a}_\lambda^{\bk}\,e^{i{\bf k\cdot x}}+{\mathrm {h.c.}}\right],
\end{eqnarray}
where the helicity vectors $\epsilon^i_\pm$ are defined so that $k_i\,  \epsilon^i_\pm=0$, $\varepsilon_{abc}\,k_b\,\epsilon^c_\pm=\mp i\,k\,\epsilon^c_\pm$, $\epsilon^i_\pm\,\epsilon^i_\mp=1$ and $\epsilon^i_\pm\,\epsilon^i_\pm=0$. Then, the functions $A_\pm$ must satisfy the equation $A_{\pm}''+(k^2 \mp k\,\chi'/f)A_{\pm}=0$.

We assume that $\chi'/a=\dot\chi\simeq\,$constant. Hence, the equation for $A_\pm$ reads
\begin{equation}
\label{d3}
\left[ \frac{d^{2}}{d\tau^{2}}+k^{2}\pm 2\,k\,\frac{\xi}{\tau} \right]A_\pm(\tau,\, k)=0\mbox{ ,}
\end{equation} 
where we have defined 
\begin{equation}
\xi\equiv\frac{\dot\chi}{2\,f\,H}\,\,,
\end{equation}
a dimensionless parameter which will determine the efficiency of the growth of the mode functions. While the quantity $\dot\chi$ is subject to the requirement that the energy in the field $\chi$ be subdominant with respect to the energy in the inflaton, the parameter $f$ is arbitrary. As a consequence, the quantity $\xi$ can take any value. As we will see, the effects under consideration are relevant for $\xi$ of the order of a few. Moreover, the approximated solution~(\ref{approx1}) below will be valid only for $\xi\gtrsim 1/4$. For these reasons, we will be interested in the case $\xi\ga{\cal {O}}\left(1\right)$.

Depending on the sign of $\xi$, one of the two modes $A_+$ or $A_-$ in (\ref{d3}) develops an instability (we assume without loss of generality that $\xi>0$, so that the mode $A_+$ will feel the instability -- remember that $\tau<0$). The other mode stays essentially in vacuum. 

The solution of~(\ref{d3}) that reduces to positive frequency for $k\,\tau\rightarrow -\infty$ is $A_\pm(\tau,\, k)=\frac{1}{\sqrt{2\,k}}[i\,F_0(\pm\xi,\,- k\,\tau)+G_0(\pm\xi,\,- k\,\tau)]$, where $F_0$ and $G_0$ are the regular and irregular Coulomb wave functions. The positive-helicity mode is rapidly amplified, and peaks at momenta $k$ for which $\left(8\,\xi\right)^{-1}\la |k\,\tau|\ll 2\,\xi$, where it is well approximated by
\begin{equation}
\label{approx1}
A_+(\tau,\, k)\simeq 
\frac{1}{\sqrt{2\,k}}\left(-\frac{k\,\tau}{2\,\xi}\right)^{1/4}\,e^{\pi\,\xi-2\,\sqrt{-2\xi \,k\,\tau}}\,.
\end{equation}

$A_+$ is thus amplified by a factor $e^{\pi\,\xi}$. On the other hand, the mode $A_-$ is not amplified by the rolling field $\chi$, and from now on we ignore it.

\subsection{Vectors generate scalar metric perturbations}

Even if the vectors are not directly coupled to the inflaton, they generate scalar perturbations through gravitational interactions. This effect was studied in~\cite{Barnaby:2012xt}, that found that the gauge invariant scalar perturbation $\hat{\zeta}$ is given by
\begin{eqnarray}\label{eq20}
\hat{\zeta}({\bf k})&&= -\frac{H^2}{4\,M_P^2}\int d\tau'\,G_k(\tau,\,\tau')\,\tau'{}^2\int\frac{d^3{{\bf q}}}{(2\pi)^{3/2}}\,\left(-1+\frac{(q-|\bk-\bq|)^2}{k^2}\right)\,\nonumber\\
&&\times\left[\hat{A}_i'({\bq},\tau')\,\hat{A}_i'({\bk}-{\bq},\tau')-\varepsilon_{iab}\,q_a\,\hat{A}_b({\bq},\tau')\,\varepsilon_{icd}\,(k_c-q_c)\,\hat{A}_d({\bk}-{\bq},\tau')\,\right]\,,
\end{eqnarray}
where we have defined the retarded Green function for the operator $d^2/d\tau^2-(2/\tau)d/d\tau+k^2$, which gives the homogeneous equation of motion of the metric perturbations
\begin{equation}
G_k(\tau,\tau')=\frac{1}{k^3\,\tau'{}^2}\left[\left(1+k^2\,\tau\,\tau'\right)\sin k\left(\tau-\tau'\right)+k\left(\tau'-\tau\right)\,\cos k\left(\tau-\tau'\right)\right] \Theta(\tau - \tau')\,. 
\end{equation}

We now compute the operator $\hat{\zeta}$ using the following approximations:
\begin{enumerate}

\item Since left-handed photons are not amplified by the rolling pseudoscalar, we consider only right-handed photons:
\begin{eqnarray}
{\hat A}_i(\tau,{\bf k})=\epsilon^i_+(\bk)\,A_+(\tau,\,\bk)\,{\hat a}_+^{\bk}+({\mathrm {h.c.}},{\bf k}\to-{\bf k});
\end{eqnarray}

\item We consider the regime where the mode function $A_+(\tau,\,\bk)$ is large and given by eq.~(\ref{approx1}). In this regime, the mode function is real;

\item We neglect the contribution from the magnetic field, the second term in brackets in eq.~(\ref{eq20}), keeping only the electric contribution, the first term in brackets, and we approximate
\begin{equation}
A_+'(\tau,\, k)\simeq 
\sqrt{-\frac{2\,\xi\,k}{\tau}}\,A_+(\tau,\, k)=\left(-\frac{k\,\xi}{2\,\tau}\right)^{1/4}\,e^{\pi\,\xi-2\,\sqrt{-2\,\xi \,k\,\tau}}\,\,.
\end{equation}
This approximation is valid for $\xi\gtrsim 1$;

\item We compute the scalar perturbations modes at late times $k\,\tau\to 0$ and we observe that the dominant contribution from the $\tau'$ integral comes from modes with $|k\,\tau'|\lesssim 1/\xi\lesssim 1$, so that
\begin{equation}
G_k(0,\tau')\simeq -\tau'/3\,;
\end{equation}

\end{enumerate}

This way the integral in $d\tau'$ can be computed explicitly, and the expression of $\hat{\zeta}$ reduces to 
\begin{equation}\label{hatzeta}
{\hat \zeta}({\bf k})=\int d^3{{\bf q}}\, F_\zeta(\bk,\,\bq)\,\hat{\cal O}_{i\, \bq}\,\hat{\cal O}_{i
, \bk-\bq}.
\end{equation}
where ${\cal O}_{l\, \bk} \equiv \epsilon^l_+(\bk)\, \left(\hat{a}_+^\bk+ \hat{a}_+^{-\bk}{}^\dagger\right)$, and where
\begin{equation}
F_{\zeta}(\bk,\,\bq)=-\frac{\Gamma(7)}{3\times 2^{12}\,(2\pi)^{3/2}}\frac{H^2}{M_P^2}\,\frac{e^{2\pi\xi}}{\xi^3}\,\frac{1}{(\sqrt{q}+\sqrt{|\bk-\bq|})^7}\,\left(-1+\frac{(q-|\bk-\bq|)^2}{k^2}\right)\,q^{1/4}\,|\bk-\bq|^{1/4}\,.
\end{equation}

\subsection{Vectors generate tensor metric perturbations}

The vectors modes discussed also source tensor metric perturbations. Since both scalar and tensor perturbations are produced gravitationally, tensors and scalars will have -- barring phase space effects -- the same amplitude. As we will see, phase space effects are however relevant: since photons are spin-one particles, tensors will be produced more efficiently than scalars.
 
Since now we focus on the tensor modes, our Ansatz for the metric is
\begin{equation}
ds^2=a^2(\tau)\,\left[-d\tau^2+\left(\delta_{ij}+h_{ij}\right)\,dx^i\,dx^j\right],
\end{equation}
with $h_i{}^i=h_{ij},_j=0$. The equation of motion for $h_{ij}$ reads
\begin{equation}\label{eqh}
h_{ij}''+2\,\frac{a'}{a}\,h_{ij}'-\Delta\,h_{ij}=\frac{2}{M_P^2}\,\Pi_{ij}{}^{lm}\,T^{EM}_{lm}
\end{equation}
where $\Pi_{ij}{}^{lm}=\Pi^i_l\,\Pi^j_m-\frac{1}{2}\Pi_{ij}\,\Pi^{lm}$ is the transverse traceless projector, with $\Pi_{ij}=\delta_{ij}-\partial_i\,\partial_j/\Delta$ and where $T^{EM}_{lm}$ represents the spatial part of the stress-energy tensor of the gauge field. $\Pi_{ij}{}^{lm}$ projects out the part of the stress energy tensor proportional to $\delta_{ij}$, leaving $T^{EM}_{ij}=-a^2\left(E_i\,E_j+B_i\,B_j\right)$. Next, we go to momentum space and we project the equation for $h_{ij}$ on left- and right-handed modes. We introduce the polarization tensors $\Pi_\pm^{ij}({\bk})$ as
\begin{equation}\label{polar}
\Pi_\pm^{ij}({\bk})=\frac{1}{\sqrt{2}}\epsilon_\mp^i({\bk})\,\epsilon_\mp^j({\bk})\,,
\end{equation}
and we define $h_\pm({\bk})=\Pi^{ij}_\pm({\bk})\,h_{ij}({\bk})$, that we promote to operators $\hat{h}_\pm$. Since $\Pi_\pm^{ij}\,\Pi_{ij}{}^{lm}=\Pi_\pm^{lm}$, and neglecting for the time being the solution of the homogeneous part of eq.~(\ref{eqh}), the expression of ${\hat h}_\pm$ can be found using the techniques of, {\em e.g.},~\cite{Barnaby:2009mc}
\begin{eqnarray}
&&{\hat h}_\pm(\bk)=-\frac{2\,H^2}{M_P^2}\int d\tau'\,G_k(\tau,\,\tau')\,\tau'{}^2\int\frac{d^3{\bq}}{(2\pi)^{3/2}}\,\Pi_\pm^{lm}({\bk})\times\\
&&\times\left[\hat{A}_l'({\bq},\tau')\,\hat{A}_m'({\bk}-{\bq},\tau')-\epsilon_{lab}\,q_a\,\hat{A}_b({\bq},\tau')\,\varepsilon_{mcd}\,(k_c-q_c)\,\hat{A}_d({\bk}-{\bq},\tau')\,\right].\nonumber
\end{eqnarray}

We can now compute the operators $\hat{h}_\pm$ using the same approximations discussed in the case of scalar perturbations and obtain 
\begin{equation}\label{eq21}
{\hat h}_\pm({\bf k})=\int d^3{{\bf q}}\, F^{lm}_\pm(\bk,\,\bq)\,\hat{\cal O}_{l\, \bq}\,\hat{\cal O}_{m \, \bk-\bq}.
\end{equation}
with
\begin{equation}\label{eq22}
F^{lm}_\pm(\bk,\,\bq)=-\frac{\Gamma(7)}{3\times 2^{9}\,(2\pi)^{3/2}}\frac{H^2}{M_P^2}\,\frac{e^{2\pi\xi}}{\xi^3}\,\frac{\Pi_\pm^{lm}({\bf k})}{(\sqrt{q}+\sqrt{|\bk-\bq|})^7}\,q^{1/4}\,|\bk-\bq|^{1/4}.
\end{equation}

\section{Two- and three-point functions}

Now that we have found the operator expression for the tensor and the scalar perturbations, we can compute the correlators we are interested in.

\subsection{Scalars}

Scalar perturbations for this model have been studied in detail in~\cite{Barnaby:2012xt}. The power spectrum can be computed directly from eq.~(\ref{hatzeta}), and for $\xi\gtrsim 3$ it can be well approximated by the following analytical expression:
\begin{equation}\label{pizeta}
P_{\zeta} = \frac{H^2}{8 \pi^2 \epsilon M_P^2} \left[1 + 3.2 \times 10^{-8} \frac{H^2 \epsilon}{ M_P^2} \frac{e^{4 \pi \xi}}{\xi^6} \right]
\end{equation}
where $\epsilon$ is the slow-roll parameter associated to the inflaton potential $V(\phi)$. Note that the contribution induced by the vectors is proportional to $e^{4\,\pi\,\xi}$, {\em i.e.,} to the fourth power of the amplitude of the vector field, since the process inducing scalar perturbations is of the type $AA\to\zeta$. The three-point function of $\zeta$ was also computed in~\cite{Barnaby:2012xt} and has an essentially equilateral shape. In the equilateral limit, $|\bk_1|=|\bk_2|=|\bk_3|=k$, we have
\begin{equation}
\langle \hat\zeta(\bk_1)\,\hat\zeta(\bk_2)\,\hat\zeta(\bk_3)\rangle_{\rm {equil}}=2.6\times 10^{-13}\,\delta(\bk_1+\bk_2+\bk_3)\,\frac{H^6}{M_P^6}\,\frac{e^{6\pi\xi}}{\xi^9}\,.
\end{equation}
%

\subsection{Tensors}

The two-point function of the tensors was first computed in~\cite{Sorbo:2011rz}. The power spectra of the left-handed and right-handed tensors can be computed directly from eq.~(\ref{eq21}), and, again for $\xi\gtrsim 3$, they can be well approximated by the following analytical expressions:
\begin{eqnarray}\label{pitensor}
P^{t,+} = \frac{H^2}{\pi^2 M_P^2} \left(1+ 8.6 \times 10^{-7} \frac{H^2}{M_P^2} \frac{e^{4 \pi \xi}}{\xi^6} \right)\\
P^{t,-} = \frac{H^2}{\pi^2 M_P^2} \left(1+ 1.8 \times 10^{-9} \frac{H^2}{M_P^2} \frac{e^{4 \pi \xi}}{\xi^6} \right)
\end{eqnarray}
where the terms $H^2/\pi\,M_P^2$ are associated to the usual amplification of vacuum fluctions of the gravitational waves in a de Sitter Universe. Note the different numerical coefficients, signaling parity violation, for the left- and the right-handed gravitons, in the $\xi$-dependent part of the spectrum. In particular, the right-handed gravitons have larger amplitude than the left-handed ones (this is a direct consequence of the fact that they are produced by right-handed photons). Inspection of the equations leading to the expressions~(\ref{pitensor}) shows that the difference in amplitude is mostly due to the process in which two almost collinear photons generate one graviton. In this regime, in fact, production of left-handed gravitons will be suppressed by helicity conservation. Also, comparison with  the last term in eq.~(\ref{pizeta}) shows that scalar perturbations are produced by vectors with an efficiency comparable to that of left-handed photons, with a  two-point function that is about $200$ times smaller than that of right-handed photons. Finally, it is worth stressing that the numbers of order $10^{-8}$ and $10^{-7}$ appearing in eqs.~(\ref{pizeta}) and~(\ref{pitensor}) do not derive from any tunable parameter, but are actual numerical quantities that originate from geometric factors such as powers of $2\,\pi$.

\begin{figure}
\centering
    \includegraphics[width=.5\textwidth]{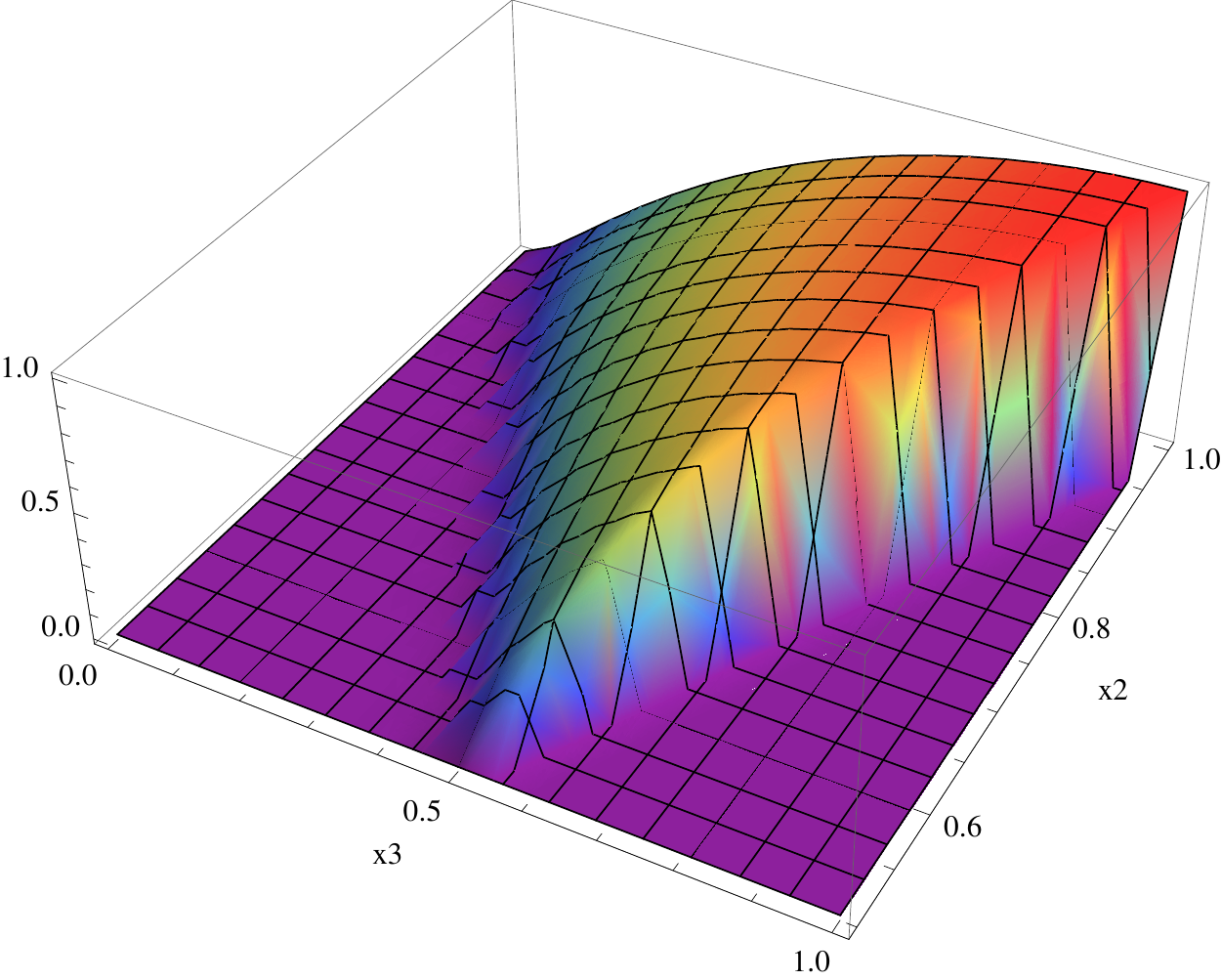}
    \caption{Plot of the shape of the three-point function~(\ref{3ptgen}) of the right handed gravitons.}
    \label{fig1}
\end{figure}

In the present paper we are interested in the three-point function of the tensors: using Wick's decomposition, the property $\langle {\cal O}_{i\, \bq}\,{\cal O}_{j\, \bk}\rangle=\delta(\bk+\bq) \epsilon_{i +}({\bf q}) \epsilon^*_{j+}(-{\bf k})$  and the property $F_\lambda(\bk,\,\bq)=F_\lambda(\bk,\,\bk-\bq)$ we obtain
\begin{align}\label{3ptgen}
\langle h_\lambda(\bk_1)&\,h_\lambda(\bk_2)\,h_\lambda(\bk_3)\rangle=4\,\delta(\bk_1+\bk_2+\bk_3)\int d^3\bq \,F_{\lambda\, ab}(\bk_1,\,\bq)\,F_{\lambda\, cd}(\bk_2,\,-\bq)\,F_{\lambda\, ef}(\bk_3,\,\bq-\bk_1) \nonumber\\
& \epsilon_{a+}({\bf q})  \epsilon_{b+}({\bf k}_1-{\bf q}) \epsilon_{c+}(-{\bf q}) \epsilon_{d+}({\bf k_2+q}) \epsilon_{e+}({\bf q-k}_1) \epsilon_{f+}({\bf k}_3+{\bf k}_1-{\bf q})+(\bk_2\leftrightarrow\bk_3)\,.
\end{align}
Since right-handed gravitons have a larger amplitude than left-handed ones, we compute this explicitly for $\lambda=+$. In Figure~\ref{fig1} we display the shape of the three-point function $\langle h_+(\bk_1)\,h_+(\bk_2)\,h_+(\bk_3)\rangle$ showing that nongaussianities  in the tensor sector are very close to equilateral, as they are in the scalar sector~\cite{Barnaby:2012xt}. This is expected from the fact that these tensors are produced by sub-horizon dynamics, so that all three modes behave identically; the integrals in $\tau'$ are dominated by  a certain value of $\tau'={\cal{O}}\left((k\,\xi)^{-1}\right)$, and since all three of the $\tau'$ integrals are the same, the three-point function is peaked when all three ${\bf k}$'s have the same magnitude.

We compute the amplitude of nongaussianities in the equilateral limit, $|\bk_1|=|\bk_2|=|\bk_3|=k$, obtaining 
\begin{equation}
\langle \hat{h}_+(\bk_1)\,\hat{h}_+(\bk_2)\,\hat{h}_+(\bk_3)\rangle_{\rm {equil}}=6.1\times 10^{-10}\,\frac{H^6}{M_P^6}\,\frac{e^{6\pi\xi}}{\xi^9}\,\frac{\delta(\bk_1+\bk_2+\bk_3)}{k^6}\,,
\end{equation}
a factor $\sim 2300$ larger than the three-point function for the scalar perturbations. The fact that $\langle h_+^3\rangle\sim {\cal O}(10^3)\langle \zeta^3\rangle$ is consistent with the fact that the component of $\langle h_+^2\rangle$ sourced by the vectors is a factor ${\cal O}(10^2)$ larger than the component of $\langle \zeta^2\rangle$ sourced by the vectors, once one observes that the three-point function scales as the two-point function to the power $3/2$.

%

\section{General effect on $\delta T$ from tensors and scalars in the flat sky approximation}

The analysis of the previous section shows that in this model tensor perturbations have larger nongaussianities than scalar ones. However, in order to make this statement precise, we have to compare the relative magnitude of the observables induced by the quantities computed in sections III. A. and III. B. To do so, we will compute the effect of both the graviton and the scalar three-point function on the observable three-point function of the temperature fluctuation in the Cosmic Microwave Background.

We will compare the relative contribution to $\delta T$ from the tensor and scalar fluctuations under some simplifying approximations: we will neglect the current acceleration of the Universe and will study the temperature fluctuations in a matter dominated Universe ({\em e.g.}, we will assume that the scalar perturbations affect the temperature fluctuations only through the Sachs-Wolfe effect); since the full calculation of the nongaussianities in the temperature fluctuations is especially challenging, we will work in the flat sky approximation~\cite{Babich:2004gb}, valid for multipoles with $\ell\gg 1$; we will consider scales that entered the horizon during matter domination ($\ell \lesssim 100$) since at smaller scales tensors are suppressed and do not give a sizable contribution to temperature fluctuations. Our analysis will therefore be valid only in the relatively narrow range $1\ll \ell\lesssim 100$, but we consider this to be sufficient to get a acceptable estimate of the size of the nongaussianities.

Let us now review the formalism of the flat sky approximation. Let $\delta T(\tau_0,\, {\bf x}_0,\,{\bf n})$ be the temperature fluctuation measured by an observer at the spatial location ${\bf x}_0$ at time $\tau_0$ looking in direction ${\bf n}$, where {\bf n} is defined to be a unit vector. We define 
\begin{equation}
a(\tau_0,\, {\bf x}_0,\,{\bf l})\equiv \int \frac{d^2{\bf n}}{2\pi}\,e^{-i{\bf l}{\bf n}}\frac{\delta T}{T}(\tau_0, {\bf x}_0,\,{\bf n})
\end{equation}
where ${\bf l}$ is a unitless two-dimensional vector satisfying $|{\bf l}|\gg 1$, and the Fourier transform
\begin{equation}
a(\tau_0,\,{\bf k},\,{\bf l})\equiv \int \frac{d^3{\bf x}_0}{(2\pi)^{3/2}}\,e^{i\,{\bf k}\,{\bf x}_0}a(\tau_0,\, {\bf x}_0,\,{\bf l})\equiv\int \frac{d^2{\bf n}}{2\pi}\,e^{-i{\bf l}{\bf n}}\frac{\delta T}{T}({\bf k},\,{\bf n})\,.
\end{equation}
Then the $C_{\bf l}$s determining the two-point function are defined as the expectation value of the correlator of $a(\tau_0,\,{\bf l})$ and $a(\tau_0,\,{\bf l}')$, that is,
\begin{eqnarray}
C_{\bf l}\,\delta^{(2)}({\bf l}+{\bf l}')&=& \int \frac{d^3{\bf k}_{1}\,d^3{\bf k}_{2}}{(2\pi)^3} \langle a(\tau_0,\, {\bf k}_1,\,{\bf l})\,a(\tau_0,\, {\bf k}_2,\,{\bf l}')\rangle\nonumber\\
&=&\int \frac{d^3\bk_1\,d^3\bk_2}{(2\pi)^3}\,\int \frac{d^2{\bf n}_1\,d^2{\bf n}_2}{(2\pi)^2}e^{-i({\bf l}{\bf n}_1+{\bf l}'{\bf n}_2)}\langle \frac{\delta T}{T}(\bk_1,\,{\bf n}_1)\,\frac{\delta T}{T}(\bk_2,\,{\bf n}_2)\rangle .
\end{eqnarray}

In an analogous way we derive the expression for the three-point function
\begin{equation}
B_{{\bf l_1},{\bf l_2},{\bf l_3}}\,\delta^{(2)}({\bf l}_1+{\bf l}_2+{\bf l}_3)=\int \frac{\Pi_{i=1}^3 d^3\bk_i}{(2\pi)^{9/2}}\,\int \frac{\Pi_{i=1}^3\,d^2{\bf n}_i}{(2\pi)^3}e^{-i\sum_{i=1}^3{\bf l}_i{\bf n}_i}\langle \frac{\delta T}{T}(\bk_1,\,{\bf n}_1)\,\frac{\delta T}{T}(\bk_2,\,{\bf n}_2)\,\frac{\delta T}{T}(\bk_3,\,{\bf n}_3)\rangle\,.
\end{equation}

\subsection{Two-point functions}

To test the validity of the approximation and to apply it to temperature fluctuations generated both by scalar and by tensor fluctuations, we first rederive the power spectrum of the temperature fluctuations induced both by scalar and by tensor perturbations in the flat sky approximation. Then we will compute the contribution of scalar and tensor perturbations to the temperature bispectrum.

\subsubsection{Sachs-Wolfe effect}

We will assume that the scalar perturbations $\hat{\zeta}$ affect the temperature fluctuations only through the Sachs-Wolfe effect, so that for scales which are not too short ({\em i.e.}, for $l \ll 200$),
\begin{align}
\frac{\delta T}{T}(\tau_0, {\bf x}_0, {\bf n})_{l \ll 200} = \frac{1}{3} \int \frac{d^3 {\bf k}}{(2 \pi)^{\frac{3}{2}}} \Phi(\tau_r, {\bf k}) e^{- i {\bf k \cdot x} }
\end{align}
where $\Phi$ is the gravitational potential. ${\bf x}$ is given by ${\bf x} = {\bf x}_0 + {\bf n}(\tau_r-\tau_0)$ which is the position of the photon at $\tau_r$, the (conformal) time of decoupling. Since $\tau_r\ll\tau_0$, the time of measurement, this simplifies to  ${\bf x} = {\bf x}_0 - {\bf n}\,\tau_0$. As a consequence we have
\begin{align}
\frac{\delta T}{T}({\bf k}, {\bf n})_{l \ll 200} = \frac{1}{3} \Phi(\tau_r, {\bf k}) e^{i {\bf k \cdot n}\tau_0 }\,.
\end{align}
We are interested in modes which entered the horizon during matter domination. In this case the relevant $\Phi$ is given by $\Phi = -\frac{3}{5} \zeta$. This gives for the temperature anisotropies:
\begin{equation}
C_{\bf l}^{\zeta,\,SW}\,\delta^{(2)}({\bf l}+{\bf l}')=\frac{1}{25}\int \frac{d^3\bk_1\,d^3\bk_2}{(2\pi)^3}\,\int \frac{d^2{\bf n}_{1}\,d^2{\bf n}_{2}}{(2\pi)^2}e^{-i({\bf l}{\bf n}_{1}+{\bf l}'{\bf n}_{2})}\langle \hat\zeta(\bk_1)\,\hat\zeta(\bk_2)\rangle\,e^{i(\bk_1\,{\bf n}_1+\bk_2\,{\bf n}_1)\tau_0}\,.
\end{equation}
Now, let us define the $(xy)$ plane as the plane where both ${\bf l}$ and ${\bf l}'$ lie. Since ${\bf l}$ and ${\bf l}'$ are ``long" vectors, the $x$ and $y$ components of ${\bf n}_1$ and ${\bf n}_2$ will contribute to the integral in $d^2{\bf n}$ only when they are much smaller than unity, otherwise the oscillating phase $\propto i\,{\bf l}\,{\bf n}$ will suppress the contribution to the integral. Therefore we can assume, remembering that the ${\bf n}_i$s are unit vectors, that ${\bf n}\simeq (n_x,\,n_y,\,1)$ with $|n_x|,\,|n_y|\ll 1$. Now the integral in $d^2{\bf n}_1\,d^2{\bf n}_2$ is straightforward:
\begin{equation}
C_{\bf l}^{\zeta,\,SW}\,\delta^{(2)}({\bf l}+{\bf l}')=\frac{1}{25}\int\frac{d^3\bk_1\,d^3\bk_2}{(2\pi)}\,\delta^{(2)}({\bf l}-{\bf k_1}^\parallel\,\tau_0)\,\delta^{(2)}({\bf l}'-{\bf k_2}^\parallel\,\tau_0)\langle \hat\zeta(\bk_1)\,\hat\zeta(\bk_2)\rangle\,e^{i\,(k_1^z+k_2^z)\tau_0}\,
\end{equation}
where $\bk^\parallel$ refers to the two-dimensional projection of $\bk$ on the $(xy)$ plane. Now we can integrate on $d\bk^\parallel_1\,d\bk^\parallel_2$, obtaining
\begin{equation}
C_{\bf l}^{\zeta,\,SW}\,\delta^{(2)}({\bf l}+{\bf l}')=\frac{1}{25\,\tau_0^4}\int \frac{dk_1^z\,dk_2^z}{2\pi}\,\langle \hat\zeta(\bk_1')\,\hat\zeta(\bk_2')\rangle\,e^{i\,(k_1^z+k_2^z)\tau_0}\,,
\end{equation}
where $\bk_1'=(l_x/\tau_0,l_y/\tau_0,\,k_1^z)$ and analogously for $\bk'_2$.

At this point we must use the explicit expression of the two-point function of the scalar perturbations. Under the assumption of a scale invariant power spectrum of amplitude ${\cal P}_\zeta$:
\begin{equation}
\langle \hat\zeta(\bk_1)\,\hat\zeta(\bk_2)\rangle=\frac{2\,\pi^2}{k_1^3}{\cal P}_\zeta\,\delta^{(3)}(\bk_1+\bk_2) \, ,
\end{equation}
we obtain the final result
\begin{equation}
C_{\bf l}^{\zeta,\,SW}=\frac{2\,\pi}{25}\,\,\frac{{\cal P}_\zeta}{{\bf l}^2} \,.
\end{equation}
This should be contrasted with the exact result for a scale invariant spectrum of tensors
\begin{equation}
C_{ l}^{\zeta,\,SW}=\frac{2\,\pi}{25}\,\,\frac{{\cal P}_\zeta}{l(l+1)} \,
\end{equation}
that shows that the flat sky approximation is accurate at better than $10\%$ for $l \gtrsim 10$.
\subsubsection{Tensors}

Next, let us compute the effect of the tensors on the two-point function of the temperature fluctuations. 
The tensor contribution to the temperature fluctuations along the direction ${\bf n}$ is, in Fourier space,
\begin{equation}
\frac{\delta T^h(\tau_0,\, {\bf x}_0,\,{\bf n})}{T}=-\frac{1}{2}\int_{\tau_r}^{\tau_0} d\tau\, \int \frac{d^3 {\bf k}}{(2 \pi)^{\frac{3}{2}}} e^{-i{\bk}({\bf x}_0 - {\bf n} (\tau_0-\tau))}\,\frac{\partial\,h_{ij}(\tau,\,{\bf k}(\tau))}{\partial\tau}\,n^i\,n^j\,,
\end{equation}
with (for modes that re-entered the horizon during matter domination)
\begin{equation}
h_{ij}(\tau,\,{\bf k}(\tau))=3\,\left(\frac{\sin(k\tau)}{(k\tau)^3}-\frac{\cos(k\tau)}{(k\tau)^2}\right)\,\hat{h}_{ij}({\bf k})\,.
\end{equation}
where $\hat{h}_{ij}({\bf k})$ is computed at the end of inflation, where we match boundary conditions with the transfer function. While it is possible in principle to compute exactly the integral in $d\tau$,  one can check numerically that the temperature fluctuation is well approximated by
\begin{equation}
\frac{\delta T^h(\tau_0,\, {\bf k},\,{\bf n})}{T}\simeq \frac{1}{2}\hat{h}_{ij}({\bf k})\,e^{i{\bk}{\bf n}\tau_0} e^{-i {\bf k \cdot x}_0} \,n^i\,n^j\,,
\end{equation}
where we have used the fact that we are looking at modes well inside our horizon, $k\gg \tau_0^{-1}$.

Now, since we will be working in the flat sky approximation, we consider the regime ${\bf n}\simeq (n_x,\,n_y,\,1)$ with $|n_x|,\,|n_y|\ll 1$, so that $\frac{\delta T({\bf k},\,{\bf n})}{T}\simeq \frac{1}{2}\hat{h}_{zz}({\bf k})\,e^{i{\bk}{\bf n}\tau_0}  e^{-i {\bf k \cdot x}_0}$. Then proceeding as in the scalar example above we get 
\begin{equation}
C_{\bf l}^{h}\,\delta^{(2)}({\bf l}+{\bf l}')=\frac{1}{4\,\tau_0^4}\int\frac{dk_1^z\,dk_2^z}{2\pi}\,\langle 
\hat{h}_{zz}(\bk_1')\,\hat{h}_{zz}(\bk_2')\rangle\,e^{i\,(k_1^z+k_2^z)\tau_0}\,,
\end{equation}
Let us assume now that the tensors are chiral, with only non vanishing positive helicity modes $\hat{h}_+$. Then
\begin{equation}
\hat{h}_{zz}(\bk)=2\,\Pi^{-}_{zz}(\bk)\,\hat{h}_+(\bk)\,,
\end{equation}
where we have used the property $2\sum_{\lambda,kl}\Pi^{-\lambda}_{ij}(\bk)\Pi^{\lambda}_{kl}(\bk)\,\hat{h}_{kl}(\bk)=\hat{h}_{ij}(\bk)$, so that
\begin{align}
\langle \hat{h}_{zz}&(\bk_1)\,\hat{h}_{zz}(\bk_2)\rangle= \nonumber\\
&= 4\,\langle \hat{h}_+(\bk_1)\,\hat{h}_+(\bk_2)\rangle\,\Pi^-_{zz}(\bk_1)\,\Pi^-_{zz}(\bk_2)=\frac{4\,\pi^2}{k_1^3}\,{\cal P}_+\,\delta^{(3)}(\bk_1+\bk_2)\,\Pi^-_{zz}(\bk_1)\,\Pi^-_{zz}(-\bk_1)\,,
\end{align}
where ${\cal P}_+$ is the power spectrum of the right-handed gravitons. Using the expression~(\ref{polar}) and the property
\begin{equation}
\epsilon_+^i(\bk)\,\epsilon_-^j(\bk)=\frac{1}{2}(\delta_{ij}-\hat{\bk}_i\,\hat{\bk}_j+i\,\epsilon_{ijk}\,\hat{\bk}_k)\,,
\end{equation}
we obtain
\begin{equation}
C_{\bf l}^{h,\,+}\,=\frac{\pi}{16}\,\frac{{\cal P}_+}{\tau_0^2}\int\frac{dk_1^z}{(k_1^z{}^2+{\bf l}^2/\tau_0^2)^{3/2}}\,\left(1-\frac{k_1^z{}^2}{k_1^z{}^2+{\bf l}^2/\tau_0^2}\right)^2=\frac{\pi}{15}\frac{{\cal P}_+}{{\bf l}^2}\,,
\end{equation}
that, upon comparison with the exact result for a scale invariant spectrum of tensors
\begin{equation}
C_{\ell}^{h,\,+}\,=\frac{\pi}{15}\frac{{\cal P}_+}{(\ell+3)\,(\ell-2)}\,,
\end{equation}
shows again that the flat sky approximation is accurate at better than $10\%$ for $\ell\gtrsim 10$.

\subsection{Three-point functions}

We are now in a position to compute the three-point functions for the temperature fluctuations induced by the scalar perturbations of section II. B and by the tensors of section II. C.

\subsubsection{$\langle \delta T^3\rangle$ induced by scalar perturbations}

For the three-point function of the temperature fluctuations induced by the scalars we have, analogously to the expression that led to the two-point function:
\begin{equation}\label{bl3scal}
B_{{\bf l}_i}^{\zeta}\,\delta^{(2)}({\bf l}_1+{\bf l}_2+{\bf l}_3)=- \frac{1}{5^3\,\tau_0^6}\int\frac{dk_1^z\,dk_2^z\,dk_3^z}{(2\pi)^{3/2}}\,\langle \hat{\zeta}(\bk_1')\,\hat{\zeta}(\bk_2')\,\hat{\zeta}(\bk_3')\rangle\,e^{i\,(k_1^z+k_2^z+k_3^z)\tau_0}\,,
\end{equation}
where $\bk_i'=(l_i^x/\tau_0,l_i^y/\tau_0,\,k_i^z)$ for $i=1,2,3$.

Let us consider the equilateral limit where $|{\bf l}_1|=|{\bf l}_2|=|{\bf l}_3|\equiv l$, since we have seen that this is the regime that dominates the three-point function. The integrals in eq.~(\ref{bl3scal}) can then be computed numerically to give:
\begin{equation}
l^4\,(B_{{\bf l}_i}^{\zeta})^{\rm {equil}}=-4.0\times 10^{-16}\frac{H^6}{M_P^6}\,\frac{e^{6\pi\xi}}{\xi^9}\,.
\end{equation}

\subsubsection{$\langle \delta T^3\rangle$ induced by tensor perturbations}

Similarly, the three-point function of the temperature fluctuations induced by the tensors is given by:
\begin{equation}\label{bl3tens}
B_{{\bf l}_i}^{h}\,\delta^{(2)}({\bf l}_1+{\bf l}_2+{\bf l}_3)=\frac{1}{8\,\tau_0^6}\int\frac{dk_1^z\,dk_2^z\,dk_3^z}{(2\pi)^{3/2}}\,\langle \hat{h}_{zz}(\bk_1')\,\hat{h}_{zz}(\bk_2')\,\hat{h}_{zz}(\bk_3')\rangle\,e^{i\,(k_1^z+k_2^z+k_3^z)\tau_0}\,,
\end{equation}
where again $\bk_i'=(l_i^x/\tau_0,l_i^y/\tau_0,\,k_i^z)$ for $i=1,2,3$.

Recalling eqs.~(\ref{eq21}) and~(\ref{eq22}), we have that $\hat{h}_{zz}(\bk)$ is given by
\begin{equation}
{\hat h}_{zz}({\bf k})=\int d^3{{\bf q}}\, F_{zz}^{ab}(\bk,\,\bq)\,\hat{\cal O}_{a,\, \bq}\,\hat{\cal O}_{b, \, \bk-\bq}.
\end{equation}
with
\begin{equation}
F_{zz}(\bk,\,\bq)=-\frac{\Gamma(7)}{3\times 2^{\frac{19}{2}}\,(2\pi)^{3/2}}\frac{H^2}{M_P^2}\,\frac{e^{2\pi\xi}}{\xi^3}\,\frac{\Pi_{zz}{}^{ab}(\bk)}{(\sqrt{q}+\sqrt{|\bk-\bq|})^7}\,q^{1/4}\,|\bk-\bq|^{1/4}.
\end{equation}
The integrals in eq.~(\ref{bl3tens}) can be computed numerically  in the equilateral limit $|{\bf l}_1|=|{\bf l}_2|=|{\bf l}_3|\equiv l$,  giving
\begin{equation}
l^4\,(B_{{\bf l}_i}^{h})^{\rm {equil}}=-1.8\times 10^{-12}\frac{H^6}{M_P^6}\,\frac{e^{6\pi\xi}}{\xi^9}\,,
\end{equation}
{\em i.e.}, a factor $\sim 4500$ larger than the contribution from the scalar perturbations.

\section{Discussion}%

\begin{figure}
\centering
    \includegraphics[width=9cm]{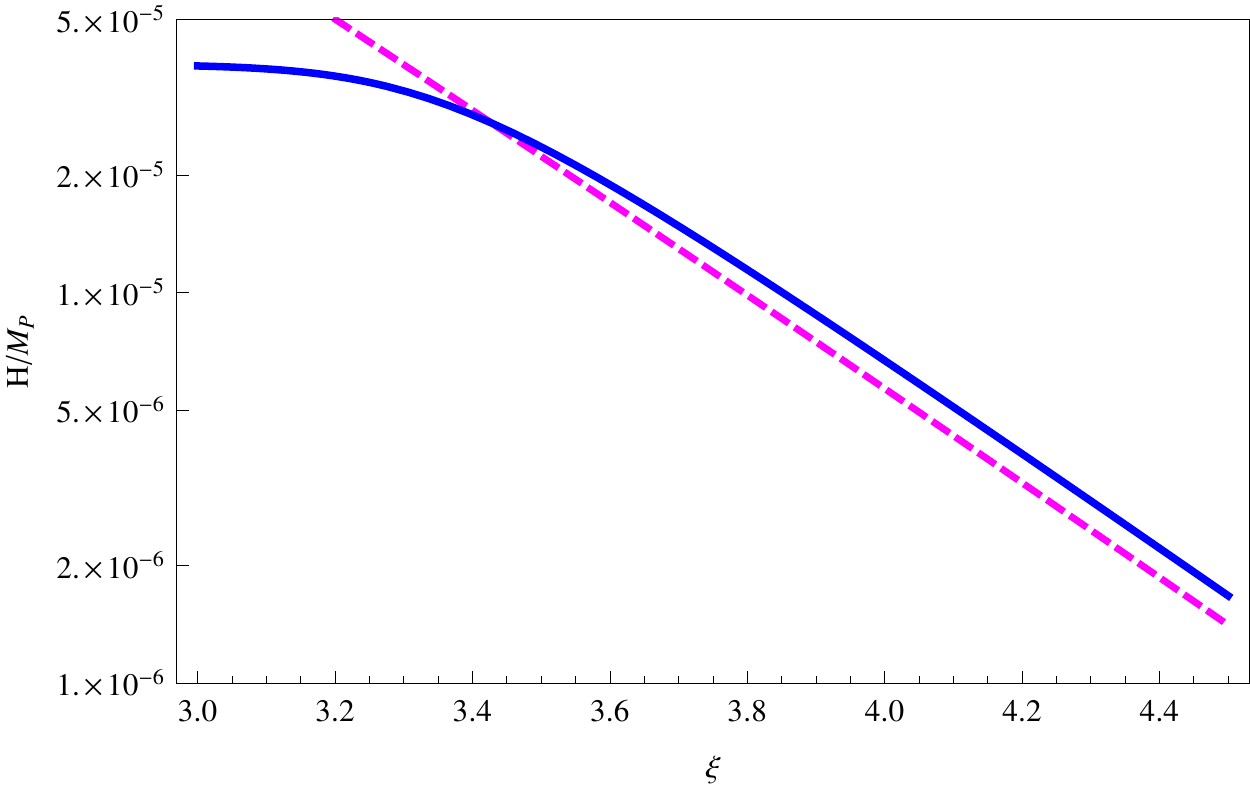}
    \caption{Maximum allowed value of $\frac{H}{M_P}$ for various values of $\xi$. The blue line uses the limit from $r< 0.11$, and the dotted pink line uses the limit from $f_{NL}^{\rm equil}<150$. The parameter space below both these lines is allowed.}
\label{fig:sub:b}
\end{figure}

In the previous sections we have computed the tensor and scalar power spectra from which we can obtain the observable $r$, the tensor to scalar ratio. We also computed the observable $B_{{\bf l}_i}^{\rm {equil}}$, the temperature bispectrum. Both depend only on the parameters $\xi$ and $H$. In Figure~\ref{fig:sub:b} we show the limits on the $(\xi,\,H/M_P)$ plane that originate from the non-observation of tensors and of nongaussianities. 

The model predicts a tensor to scalar ratio
\begin{align}
r = \frac{P^{t,+}+P^{t,-}}{P_{\zeta}} = 8.1\times 10^7\,\frac{H^2}{M_P^2} \left(1 + \frac{8.6 \times 10^{-7}}{2}\, \frac{H^2}{M_P^2}\,\frac{e^{4 \pi \xi}}{\xi^6}\right) \,.
\end{align}
The blue solid line in Figure~\ref{fig:sub:b} is obtained by applying the limit $r<0.11$ at the 95\% confidence level as published by the Planck Collaboration~\cite{Planck1}. Remarkably, in this model $r$ is not in one-to-one correspondence with $H/M_P$~\cite{Sorbo:2011rz,Cook:2011hg}, and one can have detectable tensors for arbitrarily small values of $H/M_P$. In this case the tensor spectrum would be dominated by the metric perturbations caused by the auxiliary  vector fields as opposed to the standard fluctuations caused by the inflationary expansion.

We see that for $\xi\lesssim 3.4$, where the contribution of vectors to the tensor power spectrum is weaker, the non-observation of tensors provides the strongest limit on $H/M_P$, and the the expression for the tensor power spectrum approaches the more standard expression of $P^t= \frac{2\, H^2}{\pi^2\, M_P^2}$. For these small values of $\xi$, the limit of  $r< 0.11$ translates into a limit $\frac{H}{M_P} < 3.7 \times 10^{-5}$.

To calculate $f_{NL}^{\rm equil}$, we first use the contribution to the three-point function of the temperature fluctuations from scalar perturbations
\begin{align}
\left(f_{NL}^{\rm equil}\right)_{\zeta} = 470\, \frac{H^6}{M_P^6}\,\frac{e^{ 6 \pi \xi}}{\xi^9} \, ,
\end{align}
that was first computed in~\cite{Barnaby:2012xt}. Since the three-point function of the temperature perturbations is dominated by the tensor contribution, we have
\begin{align}
f_{NL}^{\rm equil}\simeq \left(f_{NL}^{\rm equil}\right)_{h} = \frac{(B_{{\bf l}_i}^{h})^{\rm {equil}}}{(B_{{\bf l}_i}^{\zeta})^{\rm {equil}}}\,\left(f_{NL}^{\rm equil}\right)_{\zeta}=2.1 \times 10^{6}\, \frac{H^6}{M_P^6}\,\frac{e^{ 6 \pi \xi}}{\xi^9} \, .
\end{align}  
The pink dashed line in  Figure \ref{fig:sub:b} is obtained by imposing the limit $f_{NL}^{\rm equil}<150$, {\em i.e.}, twice the 68\% uncertainty published by the Planck Collaboration~\cite{Planck2}. Note that the Planck limits~\cite{Planck2} were derived for a scale invariant three-point function. However, tensors contribute to the three-point function of the temperature fluctuations only at large scales $\ell\lesssim 100$. It would be interesting to reevaluate the Planck constraints on the model taking into account the strong scale dependence of $\langle\delta T^3\rangle$. We expect the limits obtained this way would be somehow weaker than those shown in Figure~\ref{fig:sub:b}.

\begin{figure}
\centering
    \includegraphics[width=9cm]{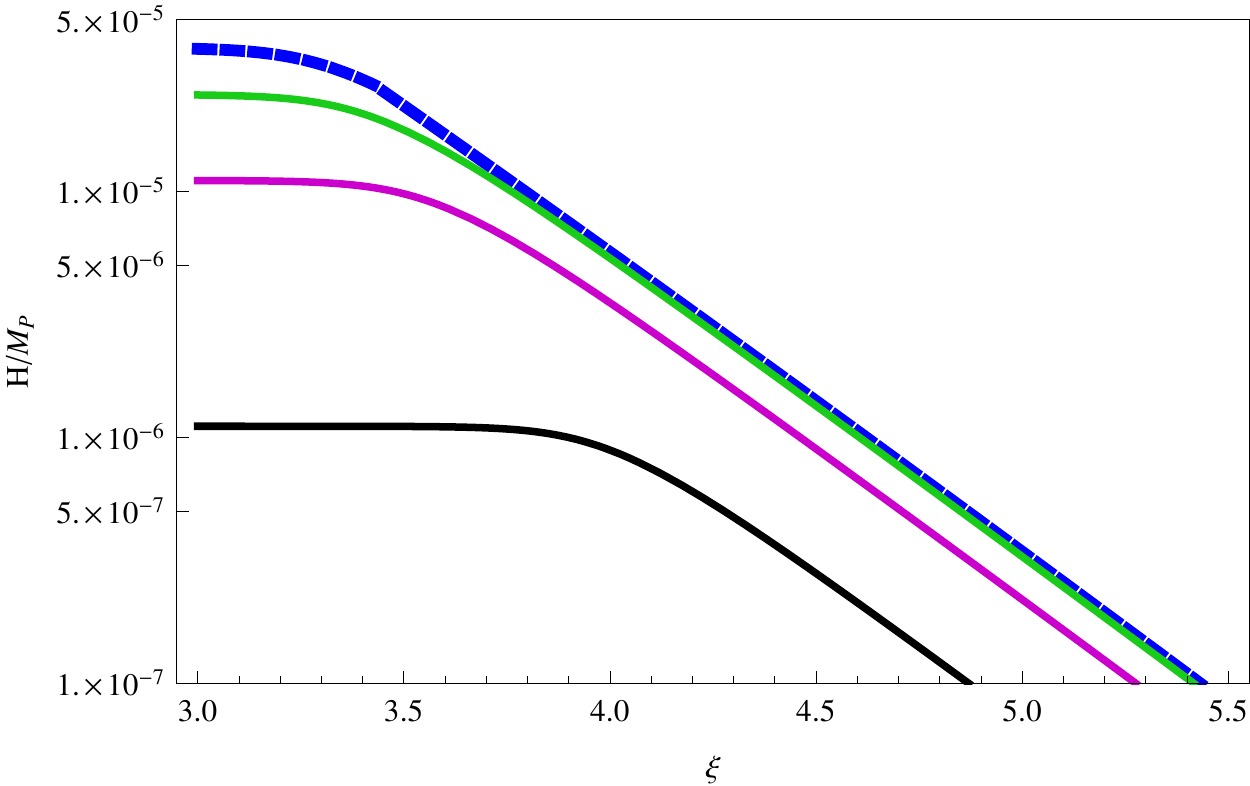}
    \caption{The figure compares the possibilities of various experiments of detecting tensor pertubations. The dotted, blue line is the maximum allowed parameters for the model using a combination of the current limits on $r$ and $f_{NL}^{\mbox{equil}}$. The solid lines correspond to the projected sensitivities to $r$ of top to bottom, Planck with polarization (green), Spider (pink), and CMBPol (black).}
\label{fig:sub:d}
\end{figure}

In Figure~\ref{fig:sub:d} we show the possibility of a detection of tensors in this model. The solid lines in the figure use the projected sensitivity of $r$ from top to bottom of: Planck, once the polarization results are complete, $r< 0.05$, Spider $r<0.01$, and a possible future CMBPol experiment $r<0.0001$~\cite{challinor}. This is in comparison to the dashed blue line which is the maximum allowed model parameters from applying the current limits on $r$ and $f_{NL}^{\rm equil}$ as displayed in Figure \ref{fig:sub:b}. We find the possibility of a detection by Spider and CMBPol for arbitrarily small $H/M_P$ for large enough $\xi$.
	
\begin{figure}
\centering
    \includegraphics[width=9cm]{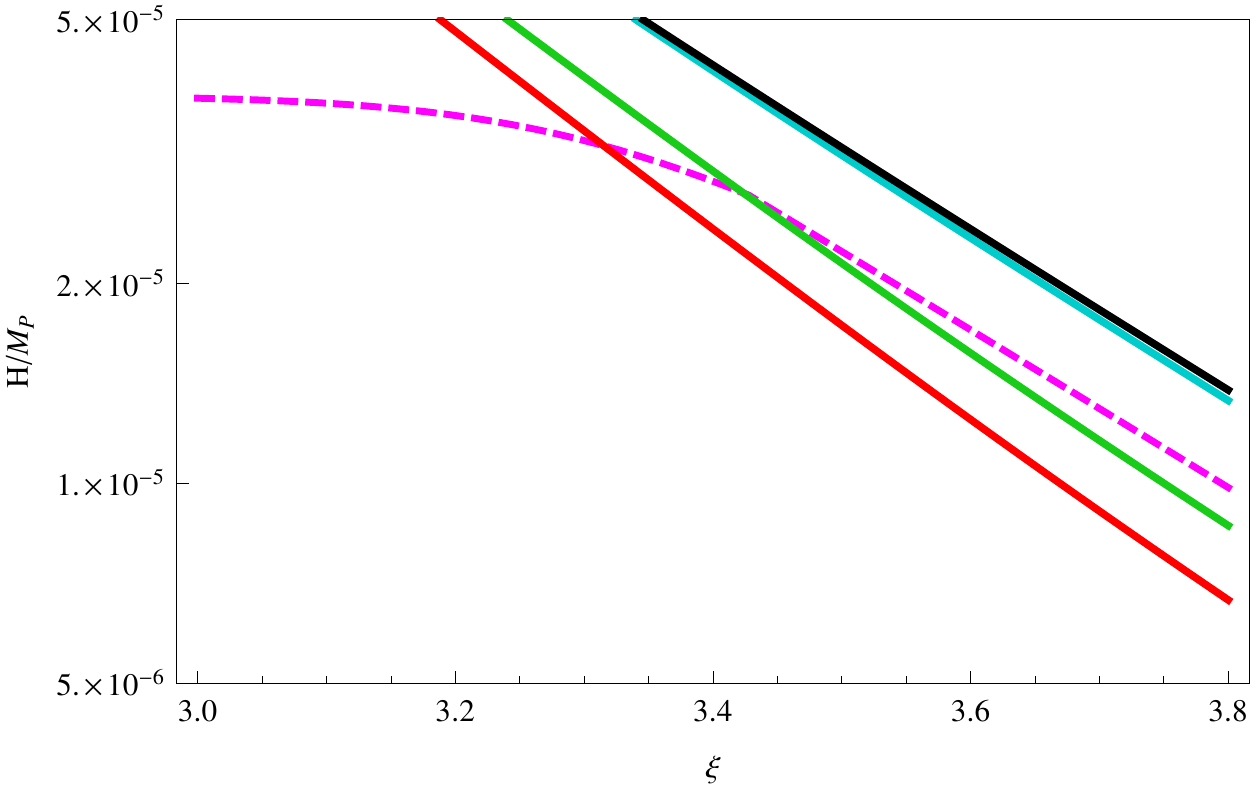}
    \caption{Comparison of detectability limits of chirality of the primordial tensors for various experiments. The dotted pink line is the maximum allowed $H/M_P$ based on current limits on $r$ and $f_{NL}^{\mbox{equil}}$. The solid lines are for $2\sigma$ detectability for the following experiments listed in order top to bottom: Planck, Spider, CMBPol, and a cosmic variance limited experiment. The experimental lines were derived from Figure 2 of~\cite{Gluscevic:2010vv}. }
\label{fig:sub:c}
\end{figure}

Next we look at the likelihood of a detection being made of the chirality of this model through a measurement of non vanishing $\langle T B\rangle$ or $\langle E B\rangle$ correlators, chirality being a more unique, interesting detection signature since it is absent from most inflationary models. We define the chirality of the tensor modes by: 
\begin{align}
\Delta \chi = \left|\frac{P_+ - P_-}{P_+ + P_-} \right| \, .
\end{align}
Using our tensor spectrum, we get:
\begin{align}
\Delta \chi = \frac{8.6 \times 10^{-7}\, \frac{H^2}{2 M_P^2} \frac{e^{6 \pi \xi}}{\xi^6}}{ 1 + 8.6 \times 10^{-7} \frac{H^2}{2 \, M_P^2} \frac{e^{6 \pi \xi}}{\xi^6} } \, .
\end{align}
 The results are shown in Figure \ref{fig:sub:c} where the pink dotted curve is the best current limit on $\xi$, using the combined limits of $r$ and $f_{NL}^{\rm equil}$ displayed in Figure \ref{fig:sub:b}. The solid curves are the $2\sigma$ detection sensitivities for various experiments, such that the parameters in the model will need to fall above these curves to allow detection by these various experiments. We use twice the detection limits published in~\cite{Gluscevic:2010vv}, to require a $2\sigma$, rather than a $1\sigma$, detection of primordial parity-violation in the CMB. We find that there is no allowed region of detection from Planck or Spider. On the other hand, a detection by a cosmic variance limited experiment or a CMBPol like experiment is allowed throughout a part of parameter space. The larger $\xi$ is, the more the tensor spectrum is dominated by the auxiliary model fields as opposed to the standard perturbations from expansion, and since it is the contribution from these auxiliary fields that violate parity, the larger $\xi$, the larger $\Delta \chi$. 

Let us note that the constraints on the system can be studied analytically. To simplify our notation, let us introduce the quantities ${\cal P}\equiv \frac{H^2}{8\pi^2\epsilon\,M_P^2}$~\cite{Barnaby:2012xt} and $X\equiv \epsilon\,\frac{e^{2\pi\xi}}{\xi^3}$, where $\epsilon$ is the slow roll parameter. Then we can write 
\begin{eqnarray}
&&{\cal P}_\zeta={\cal P}\left(1+2.5\times 10^{-6}\,{\cal P}\,X^2\right)\,\label{pzeta}\\
&&r=\frac{16\,\epsilon+5.4\times  10^{-4}\,{\cal P}\,X^2}{1+2.5\times 10^{-6}\,{\cal P}\,X^2}\,\\
&&f_{NL}^{\rm {equil}}= 1.1\times 10^{12}\,{\cal P}^3\,X^3\,,
\end{eqnarray}
where ${\cal P}_\zeta=2.5\times 10^{-9}$ is the observed value of the scalar power spectrum. Now let us assume that the second term in brackets in eq.~(\ref{pzeta}) is negligible with respect to $1$. We will check in a second that this has to be the case. Then we have ${\cal P}={\cal P}_\zeta$, and the $95\%$ Planck constraint $f_{NL}^{\rm {equil}}<150$ turns into the constraint
\begin{equation}\label{limitX}
X<2.1\times 10^5\,,
\end{equation}
from which we see that the second term in brackets in eq.~(\ref{pzeta}) has to be smaller than $2.6\times 10^{-4}$ that is indeed much smaller than unity. This implies that $r=16\,\epsilon+5.4\times  10^{-4}\,{\cal P}\,X^2$, where the second term is constrained by nongaussianities to be smaller than $0.057$. As a consequence, if $X$ saturates the bound~(\ref{limitX}) and $16\,\epsilon\ll 0.057$, then it will still be possible to detect chiral tensors in the CMB.

\section{Conclusions}%

The coupling~(\ref{int}) leads to a rich phenomenology in the tensor sector: not only can it give an observable spectrum of gravitational waves even for low values of the Hubble parameter, but it can also produce detectable $\langle T B\rangle$ or $\langle E B\rangle$ correlators, which is rare in inflationary models.  The model was first studied in the context where the field sourcing the vectors was the inflaton, in which case the main signature was in the form of scalar non-Gaussianities, leaving detectable effects in the tensor sector only in special cases, such as the coupling of the inflaton to multiple gauge fields,  or curvaton models so that the field sourcing curvature perturbations, the curvaton, is not the field which is directly coupled to the vectors, the inflaton~\cite{Sorbo:2011rz}. Another possibility is the case where the detectable signal is only possible in much larger frequency modes probed at direct detection experiments \cite{Cook:2011hg}.  Reference~\cite{Barnaby:2012xt} suggested another possibility of weakening the scalar perturbations relative to the tensors, having another scalar field other than the inflaton couple to the vectors. This is the model discussed in this paper. 

Our main result is the proof that this model provides the first example, to our knowledge, of a scenario where the three-point function of the tensors is significantly larger than the three-point function of the scalars. We have calculated the three-point function of the temperature perturbations induced both by scalar and by tensor metric perturbations, extending to primordial tensors the formalism of the flat sky approximation. This has allowed us to calculate $f_{NL}^{\rm equil}$ induced by the tensors, which sets the strongest constraints on the model for much of the parameter space. Despite these constraints, it will still be possible to detect parity-violating correlation functions in the CMB by a future CMBpol-like experiment, while such a detection by Planck and Spider will not be possible. Such a detection would appear unique compared to other inflationary model tensor signals in that it gives a non-Gaussian tensor power spectrum and a non-vanishing $\langle T B\rangle$ and $\langle E B\rangle$ correlator. 

More in general, although clearly $\langle TT \rangle$ is dominated by scalar signals, it is possible that $\langle TTT \rangle$ is dominated by tensor signals. In general this requires a mechanism generating perturbations on CMB scales which is highly non-Gaussian (true generically of particle production models among others), and which produces stronger tensor metric perturbations than scalar ones. This works out in the case of our particular model because 1. there is no direct coupling of produced vectors to the inflaton to enhance the scalar perturbations and 2. this mechanism naturally produces a larger tensor signal than scalar signal because of the phase space available for the decay. The model has relativistic vectors which are decaying into scalar and tensor perturbations, where there is naturally a larger phase space available for the decay into tensors.  Since most inflationary models produce very small non-Gaussianities, a mechanism which is swamped by other signals in the temperature two-point function could still dominate the three-point function. 

We expect the fact that tensor nongaussianities are much larger than scalar ones to have several phenomenological consequences that we did not analyze in this paper. First, since tensors affect temperature perturbations only at large scales, we expect a strongly scale dependent temperature three-point function, that should rapidly fall to very small values for $\ell\gtrsim 100$. Second, we expect a peculiar pattern of polarization bispectra. Third, since in this model tensors are chiral, we expect that the coefficient $B_{\ell_1,\ell_2,\ell_3}$ vanish for $\ell_1+\ell_2+\ell_3=$even, differently from the usual case where temperature nongaussianities are not generated by a parity-violating source. We hope to come back to these points in a forthcoming publication.

\smallskip

{\bf Note added.} After the submission of the present paper, ref.~\cite{Shiraishi:2013kxa} appeared, that performed a detailed analysis (beyond the flat sky approximation) of the CMB temperature and polarization bispectra for the system~(\ref{lagr}). In particular, the temperature bispectrum in the equilateral limit agrees well, for $\ell\lesssim 80$, with our flat sky analysis, supporting the validity of our estimate. The limit on $X$ derived from the Planck temperature bispectrum is weaker by a factor $\sim 3$ than our estimate of section 5. We attribute this discrepancy to the fact that our estimate was based on Planck constraints, that assumes a scale invariant bispectrum, while the temperature bispectrum induced by tensors is significant only at $\ell\lesssim 100$, as can be seen by comparing the red and the cyan lines in the top left panel of figure 1 in~\cite{Shiraishi:2013kxa}.

\begin{center}
{\bf Acknowledgements}\\
\end{center}

It is a pleasure to thank Chiara Caprini, Guido D'Amico, David Langlois, Ryo Namba, Marco Peloso and Bartjan Van Tent for very interesting discussions. LS thanks the institute AstroParticule et Cosmologie at the University Paris 7 for hospitality during the completion of this work. This work is partially supported by the U.S. National Science Foundation grant PHY-0855119

\end{document}